\documentclass[12pt]{article}

\usepackage{newtxtext,newtxmath}

\usepackage{graphicx}

\usepackage[letterpaper,margin=1in]{geometry}

\renewenvironment{abstract}
	{\quotation}
	{\endquotation}

\date{}

\makeatletter
\renewcommand{\fnum@figure}{\textbf{Figure \thefigure}}
\renewcommand{\fnum@table}{\textbf{Table \thetable}}
\makeatother

\usepackage{scicite}
\usepackage{braket}
\usepackage{verbatim}
\usepackage{url}

\def\scititle{
	Nanoscale imaging of ferromagnetic vortex dynamics with scanning  NV magnetometry
}

\title{\bfseries \boldmath \scititle}

\author{
	Jeffrey~Rable$^{1,2\ast}$,
	Jyotirmay~Dwivedi$^{3}$,
	Nitin~Samarth$^{3,4}$,\and Paul~Stevenson$^{1,2}$, Arun~Bansil$^{1,2}$, Swastik~Kar$^{1,2,5\ast}$\and
	\small$^{1}$Department of Physics, Northeastern University, Boston, MA 02115 \and
	\small$^{2}$Quantum Materials and Sensing Institute, Northeastern University, Burlington, MA, 01803 \and
	\small$^{3}$Department of Physics, Pennsylvania State University, State College, PA, 16802  \and
    \small$^{4}$ Q-NEXT, Argonne National Laboratory, Lemont, IL 60439 \and
 	\small$^{5}$Department of Chemical Engineering, Northeastern University, Boston, MA 02115 \and
	\small$^\ast$Corresponding author. Email: j.rable@northeastern.edu, s.kar@northeastern.edu\and
}

\begin{document} 

\maketitle

\begin{abstract} \bfseries \boldmath 
The generation and manipulation of spin waves at the nanoscale via magnetic vortices are of considerable importance because of their broad applications across magnonic and quantum technologies. Previously, fixed nitrogen-vacancy (NV) centers in diamond have been used to locally characterize vortex dynamics, and scanning NV magnetometry (SNVM) has been used to image vortices’ static stray fields. Here, we demonstrate SNVM imaging of both the static and microwave fields generated by vortices in mesoscopic permalloy structures with $\sim$50 nm spatial resolution, achieving excellent agreement with micromagnetic simulations, while revealing the effects of disorder. We further demonstrate a 40$\times$ microwave field enhancement near a vortex core and image the disorder-dependent, spatially varying, evanescent decay of these microwaves. Our ambient, tabletop technique surpasses diffraction-limited techniques' resolutions by at least 5$\times$, with far greater accessibility and throughput than synchrotron radiation-based techniques, offering new opportunities in the study and development of magnonic devices. 
\end{abstract}

\subsection*{Introduction}
\noindent
After over half a century of exponential growth in computing power, as predicted by Moore's law, technology has begun to reach the limits of transistor density in electronic integrated circuits. There are a multitude of alternative methods beyond just increasing transistor density being proposed to maintain rapid growth in computation, ranging from the high level - such as improving system architecture, interconnects, and algorithms, an approach currently taken in industry\cite{leiserson_theres_2020} - to the fundamental, where one eschews electronic current-based logic in favor of an entirely new information carrier. One such alternative is magnon spintronics, or magnonics. In magnonics, one utilizes spin waves, the excitation of magnetic order, as an information carrier, with magnons being their quasiparticle. Utilizing spin excitation, rather than the physical motion of electrons, comes with multiple potential advantages over existing CMOS systems: (i) no Joule heating, which limits the efficiency, clock speed, and compactification of electronics, (ii) data processing in the GHz to THz regimes, matching or drastically exceeding existing systems, and (iii) the ability to directly integrate non-volatile magnetic memory into data processing, avoiding interconnect bottlenecks\cite{yu_magnetic_2021,nizet_perspective_2025, flebus_2024_2024,balinskyy_magnonic_2024}. Furthermore, magnonic devices have potential to be integrated into other computational paradigms, such as quantum computing, where magnonic devices could be used to control qubits or act as a quantum bus, enabling long distance coupling between them\cite{Candido2020,fukami2021,fukami_magnon-mediated_2024,bejarano_parametric_2024}.
Short wavelength spin waves are particularly critical for the creation of fast, miniaturized devices. One potential method of producing and manipulating these spin waves is through magnetic texture dynamics; nanoscale ferromagnetic textures possess dynamics in the MHz to GHz range, are highly stable, and can be manipulated with magnons, DC currents, AC currents, and spin transfer torques\cite{han_mutual_2019,gao_interplay_2023,chang_spin_2020,carolin_behncke_spin-wave_2018,dieterle_coherent_2019,koraltan_steerable_2024,mayr_spin-wave_2021,van_de_wiele_tunable_2016,wintz_magnetic_2016,hamalainen_tunable_2017}. In particular, ferromagnetic vortices have extremely rich linear and nonlinear dynamics - they possess both a gyrating mode and magnon modes\cite{park_interactions_2005,buess_excitations_2005,Trimble2021,neudecker_modal_2006}, nonlinear dynamics from the hybridization of these modes\cite{korber_nonlocal_2020,heins_self-induced_2026,devolder_time-resolved_2025}, and frequency multiplication\cite{wang_spin-wave_2026,trimble_parallel_2024}. Furthermore, they have been demonstrated to be versatile, wide bandwidth, short wavelength spin wave emitters\cite{wintz_magnetic_2016,carolin_behncke_spin-wave_2018,koraltan_steerable_2024,chang_spin_2020}.
However, it is incredibly difficult to image and characterize short wavelength spin waves, both texture generated and otherwise, with high spatial resolution. The most accessible methods used to probe these dynamics - bulk electric and optical measurements, such as time resolved magneto-optic Kerr effect (TR-MOKE) - fail to fully image and characterize the nanoscale dynamics below the diffraction limit, making it more difficult to directly measure magnon wavelength and other important features of these modes\cite{chang_spin_2020}. Conversely, time resolved scanning X-ray transmission microscopy (TR-STXM), an extremely powerful characterization technique capable of achieving up to single-nm spatial resolutions\cite{feggeler_scanning_2023}, has been used to great success\cite{wintz_magnetic_2016,carolin_behncke_spin-wave_2018,koraltan_steerable_2024,mayr_spin-wave_2021}, but requires synchrotron access, placing restrictions on sample substrates, accessibility, and characterization throughput. 

One potential alternative tabletop measurement technique is scanning nitrogen-vacancy center magnetometry (SNVM). SNVM has a lateral resolution only limited to the NV-sample separation\cite{xu_minimizing_2025,Broadway2020}, providing an advantage over diffraction limited optical techniques such as TR-MOKE and Brillouin Light Scattering, and can quantitatively characterize both DC fields and AC magnetic fields down to the 100's of $nT / \sqrt{Hz}$\cite{sun_magnetic_2021,zhang_ac_2021,appel_nanoscale_2015}. Previously, SNVM has allowed for precise DC imaging of vortex textures, establishing it as one of the few measurement techniques capable of resolving the core\cite{tetienne_quantitative_2013,rondin_stray-field_2013,sfeir_room_2025}, while fixed NV magnetometry has been used to characterize MW enhancement near a proximal vortex\cite{Wolf2016}, vortex gyrating dynamics\cite{Trimble2021}, stochastic vortex dynamics\cite{Badea2018}, vortex-mediated parallel pumping of magnons\cite{trimble_parallel_2024}, and vortex induced harmonic generation\cite{wang_spin-wave_2026}. Further improving our understanding of the wide spectrum of vortex dynamics has the potential to not only improve spintronic technologies, but quantum technologies as well - both the static and MW fields generated by both saturated and vortex-containing magnetic discs are applicable to quantum control schemes with solid state qubits. Their applications in quantum magnonics include using their high-gradient stray fields to improve the addressibility of neighboring spin qubits\cite{Wolf2016,Badea2018}, to couple distant qubits using magnon transduction\cite{Candido2020,bejarano_parametric_2024}, locally drive qubits\cite{Badea2023}, or mediate local interactions between spins for computation and sensing\cite{wolf_strong_2017}. Here, primary difficulties for implementation include decreases in coherence and gate fidelity from both thermal and driven magnetic dynamics\cite{Du2017,Badea2023}, and positioning issues, where optimal local control depends on precise qubit placement relative to the desired feature. Through scanning NV measurements, one can characterize numerous device properties and their impact on qubits with high spatial resolution, allowing for rapid prototyping prior to integration with fixed qubits. 

Here, using quantitative and semi-quantitative measurement techniques, we directly image both the static and microwave fields generated by vortex magnon modes in two permalloy (NiFe alloy, Py) features, a 1 \textmu m square and a 6 \textmu m diameter disc, using SNVM, shown in Fig.~\ref{fig:2PeakODMRSquare}(A). We demonstrate that this technique is viable for imaging multiple different NV-resonant magnon modes - wall modes confined to an extended vortex structure in the 20 nm thick, Py square, and an azimuthal mode concentrated near the vortex core in the 20 nm thick disc. Through optically detected magnetic resonance (ODMR) measurements, we obtain high resolution quantitative images of the features' static magnetic textures and semi-quantitative images of their GHz dynamics in a small external field. We also quantitatively image these MW fields at a lower resolution using Rabi oscillation scans, and measure both the power and height dependence of the MW fields near the vortex with Rabi oscillation measurements at specific sites. In the square feature, we positively identify and characterize the vortex structure through both our static and dynamics measurements, showing the applicability of the technique on a prototypical extended vortex structure that is robust against disorder. In the disc feature, which is less robust against disorder\cite{sfeir_room_2025}, a vortex texture cannot be definitively identified through the static stray field map, but the presence of an azimuthal vortex mode, localizing the core location, becomes evident in the dynamical measurements when comparing to micromagnetic simulations. Additionally, we can characterize both the power dependence and evanescent decay of the AC fields generated by these modes, revealing a strong field enhancement and highly position dependent, short-range exponential decay. 



\subsection*{Results}

\begin{figure} 
	\centering 
	\includegraphics[width=\textwidth]{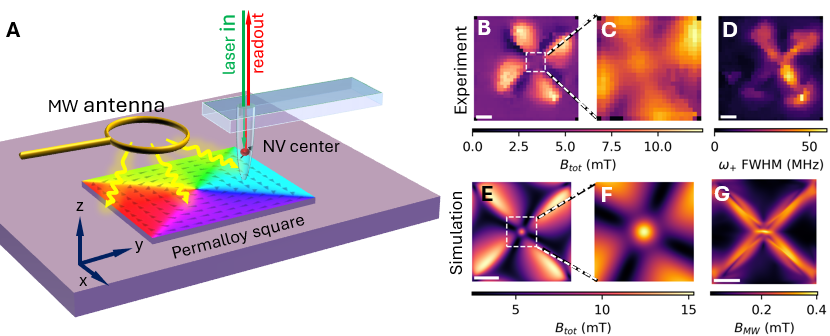}
	\caption{\textbf{Schematic of the experimental setup and primary results on a 1 \textmu m square.} Scale bars are 250 nm.
		(A) A diagram of our measurement scheme. The permalloy disc contains a vortex, and a manually positioned microwave antenna drives both our ferromagnetic dynamics and NV spin transitions. Our NV-containing diamond tip is rastered across the sample in the xy direction during a scan, and lifted measurements are performed by increasing tip height in z. (B) Total magnetic field of a 1 \textmu m square measured using ODMR showing a vortex with Néel domain walls extending towards the corners. The background field is approximately 4 mT, and the central vortex core, which spanned only one pixel, was filtered out during our image processing (see Supplementary Materials). (C) Magnetic field of a 300 nm by 300 nm area around the vortex core; the size of the field from the core reflects our 54 nm resolution. (D) FWHM of the ODMR peak at the $\ket{0}\rightarrow\ket{1}$ ground state NV transition frequency, showing power broadening from the microwave response of the square. (E) Total simulated magnetic field in a vortex-containing Py square. (F) 300 by 300 nm image of the simulated vortex core stray field. (G) Simulated microwave magnetic field produced by the square under a 0.1 mT amplitude, 2.85 GHz excitation showing a strong response at the core and along the domain walls.}
	\label{fig:2PeakODMRSquare} 
\end{figure}

The results of an ODMR scan measuring the static magnetic field over the whole 1 \textmu m square in the presence of a 4.3 mT, out-of-plane bias field are shown in Fig.~\ref{fig:2PeakODMRSquare}(B-C). Here, from measurements of both the $\ket{0}\rightarrow\ket{-1}$ and $\ket{0}\rightarrow\ket{1}$ ground state spin transitions, which are Zeeman shifted in the presence of magnetic fields, we extracted the static stray fields both along and perpendicular to the NV center's quantization axis and computed the total field\cite{VanderSar2015,Dovzhenko2018}. The results in Fig.~\ref{fig:2PeakODMRSquare}(B) show the same stray field configuration as the corresponding micromagnetic simulations (Figs.~\ref{fig:2PeakODMRSquare}(E)) and previous works in the literature\cite{breitenstein_stray_2008,rondin_stray-field_2013,sfeir_room_2025} - an extended, pinwheel-like structure composed of domain walls pinned by the corners of the square. Measuring a smaller 300 x 300 nm area around the center of the square in Fig.~\ref{fig:2PeakODMRSquare}(C) and (F), we can directly identify the central vortex core in both experiment and simulation. The spatial resolution of these measurements - the minimum separation needed to resolve individual features - is limited by our 54 nm NV-sample separation (see Supplementary Materials), and the simulation resolution is limited to 50 nm by the height where we solved for the stray field; while the core magnetization spans approximately 10-15 nm, its stray field covers a significantly larger area at our standoff distance. Fourier decomposition of the stray field into its cartesian components and comparison to micromagnetic simulations is consistent with a counterclockwise vortex circulation (see Supplementary Materials), and the structure of the core is consistent with a positive polarity\cite{rondin_stray-field_2013}.

Previous works have used ODMR contrast to image magnon modes in saturated single crystal materials with great success\cite{Simon2022,Zhou2021,simon_directional_2021}; here, we extend those capabilities to polycrystalline, textured materials using a combination of ODMR and pulsed measurements. Because ODMR contrast can saturate and is dependent on stray magnetic field as well as microwave field amplitude, it fails to accurately capture dynamics in these systems. Instead, we use ODMR full width at half maximum (FWHM), which, in the power-broadened regime (approximately $\geq 2$ MHz Rabi frequency), is equal to $\frac{\gamma_{\text{NV}} B_{\text{MW}}\sqrt{\beta}}{2\sqrt{2}}$, where $\gamma_{\text{NV}}$ is the NV center's gyromagnetic ratio, $B_{\text{MW}}$ is the amplitude of the MW field perpendicular to the NV center's quantization axis, and $\beta$ is a constant dependent on optical and spin initialization rates\cite{Dreau2011,Wolf2016}. The FWHM of the $\ket{0}\rightarrow\ket{+1}$ ODMR peak from our measurements of the square can be seen in Fig.~\ref{fig:2PeakODMRSquare}(D), and the micromagnetic simulations of the square dynamics under a 0.1 mT, NV-resonant driving field can be seen in Fig.~\ref{fig:2PeakODMRSquare}(G). In both our simulations and experiment, we find regions of large MW fields (FWHM) radiating outwards from the core along the domain walls extending to the corners of the square, showing dynamics predominantly along the extended structure of the vortex, rather than at the core itself. The primary difference between these two results - the anisotropy in the experimentally measured dynamics - could arise from inexact simulations, as we did not consider any disorder in the film and estimated the applied MW field direction. While these wall modes have been previously reported in the literature\cite{park_imaging_2003, stoll_high-resolution_2004} in significantly larger square features, we demonstrate a nearly 10$\times$ higher measurement spatial resolution than comparable tabletop measurements. The significantly higher wall mode frequency we observed compared to previous experimental works\cite{park_imaging_2003,stoll_high-resolution_2004,bailleul_microwave_2007} is likely because of the significantly smaller size of the square we are measuring on, which will result in stronger pinning of the texture. 

\begin{figure} 
	\centering 
	\includegraphics[]{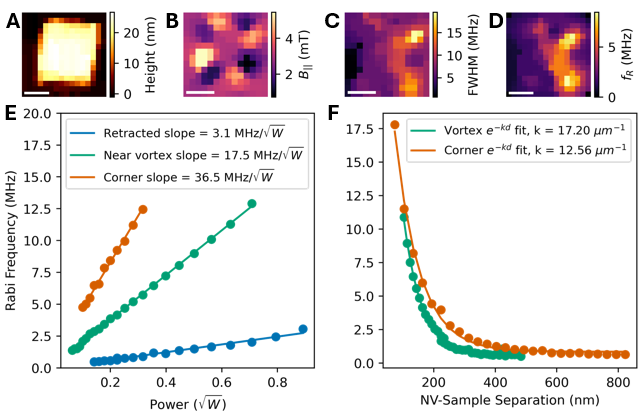} 
	\caption{\textbf{Rabi oscillation measurement results.} Scale bars are 500 nm. 
		(A) Topography of the permalloy square. (B) ODMR measurement results on the permalloy square giving the approximate on-axis static field at 125 nm lift.. (C) FWHM results extracted from the ODMR measurement shown in (B), showing qualitative variation in MW field strength. (D) Rabi oscillation scan results at 125 nm lift showing the quantitative variation in microwave field strength across the disc. These results correspond well with our ODMR FWHM. (E) Power dependence of Rabi oscillations measured with the tip retracted off the sample, positioned near the vortex core, and near the bottom right corner of the square. We find a 5.5$\times$ increase in MW power in the near-core measurement over the retracted measurement and a 11$\times$ increase near the corner. (F) Measurement of the evanescent decay of the enhanced fields at the core and the corner.}
	\label{fig:RabiSquare} 
\end{figure}

To quantify the magnitude of the GHz AC field, we subsequently performed Rabi oscillation measurements. In these scans, we measure the height (Fig.~\ref{fig:RabiSquare}(A)) with a linescan, and subsequently lift the tip an additional 125 nm, giving a total NV-sample separation of 179 nm, improving readout efficiency. Then, at each point, we perform a series of ODMR measurements, which provide the NV transition $\ket{0}\rightarrow\ket{1}$ frequency, and Rabi oscillation measurements, where we coherently drive the NV spin between the $\ket{0}$ and $\ket{1}$ states. The Rabi oscillation frequency, $f_R$, is equal to $\frac{\gamma_{\text{NV}} B_{\text{MW}}}{2\pi\sqrt{2}}$\cite{Wolf2016}, giving a fully quantitative measurement of the AC field. The ODMR results, shown in Fig.~\ref{fig:RabiSquare}(B) and Fig.~\ref{fig:RabiSquare}(C), match our previous results, where we see a vortex in the static stray field measurements and an anisotropic MW field in the disc radiating out from the vortex core. Our Rabi oscillation measurements in Fig.~\ref{fig:RabiSquare}(D) show a similar result - an anisotropic field emanating from the core, with a substantial increase in the measured microwave field. Furthermore, the ODMR FWHM and Rabi frequencies strongly correlate with one another and lie predominately in the power-broadened regime (see Supplementary Materials), confirming the validity of our approach to rapid, high resolution measurements of magnetic dynamics in textured systems.

We can find the total MW field enhancement produced by the vortex by measuring the power dependence of Rabi oscillations with the tip positioned over the vortex core and comparing to the retracted tip, driven solely by our MW antenna. The results of these measurements, which were done without lift, are shown in Fig.~\ref{fig:RabiSquare}(E); they demonstrate an almost 5.5$\times$ increase in the MW field amplitude near the vortex compared to the retracted case and an almost 12$\times$ increase at the corner. This demonstrates that these wall modes offer relatively large MW field enhancements, comparable to previous measurements near the core in vortex-containing discs\cite{Wolf2016}, though we also note that our $\geq$ 50 nm standoff distance is over twice as high as these previous results, indicating that even larger field enhancements are possible. Similarly, performing Rabi oscillation measurements at different heights over the sample, we can measure the decay of the evanescent waves produced by wall modes. Because these Rabi frequencies directly correlate to evanescent wave amplitude, their height dependence can then be fit with an exponential function, $Ae^{-kd} + f_{Ro}$, where $A$ is the amplitude of the Rabi frequency increase at the surface of the disc, $k$ is the decay constant of the evanescent field, $d$ is the NV-sample separation, and $f_{Ro}$ is the Rabi frequency of the retracted tip, driven solely by our MW antenna. We find that they decay extremely rapidly, with $k = 17.20 \mu m^{-1}$ (58.14 nm decay length) at the core and $k = 12.56$ \textmu m$^{-1}$ (79.62 nm decay length) at the corner of the square, which are consistent with micromagnetics (see Supplementary Materials). 

\begin{figure} 
	\centering
	\includegraphics[]{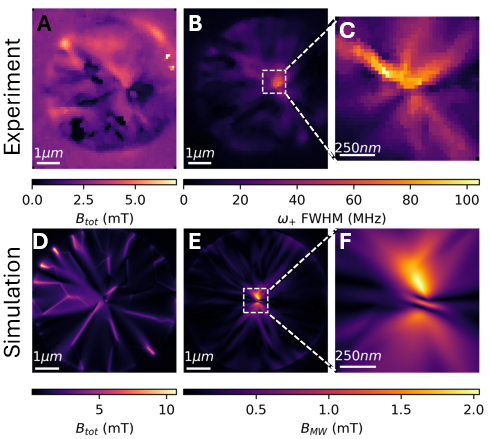}
	\caption{\textbf{Measurements on vortex-containing, disordered disc}
		 (A) Total magnetic field measured using ODMR. The background field is approximately 2.1 mT. (B) FWHM of the ODMR peak at the $\ket{0}\rightarrow\ket{1}$ ground state NV transition frequency, showing power broadening from the microwave response of the disc. (C) High resolution image of the ODMR FWHM at the vortex core. (D) Total simulated stray field 50 nm above a permalloy disc with 1 \textmu m grains and a vortex texture. (E) Simulated microwave magnetic field from the disc under a 0.1 mT amplitude, 2.85 GHz excitation. (F) High resolution image of the microwave field near the simulated vortex core.}
	\label{fig:2PeakODMRDisc} 
\end{figure}

To study how vortex modes evolve in the absence of strong geometric pinning sites (corners) and the presence of static disorder, we measure a vortex-containing 6 \textmu m Py disc, which possesses a 100's of MHz-wide azimuthal mode overlapping the NV transition frequencies. Previous reports of NV center measurements on similar discs suggest that it becomes substantially harder to determine the exact structure of the magnetization compared to an ideal disc or a square feature; the disorder leads to stray fields throughout the feature\cite{sfeir_room_2025,rondin_stray-field_2013,wang_spin-wave_2026}. Our ODMR results on this feature, shown in Fig~\ref{fig:2PeakODMRDisc}(A), show a complex texture radiating throughout the disc, unlike with an ideal vortex, where one would expect to only see a localized stray field from the core (see Supplementary Materials). Comparing to our simulation results of a polycrystalline, vortex-containing disc with  1 \textmu m grains in Figs.~\ref{fig:2PeakODMRDisc}(E), we find qualitative similarities - large fields around the edges of the disc and near the disc center - but cannot definitively identify a vortex from the static field alone. Even in our simulated stray fields, it is not apparent that there is a vortex core because the stray fields from the disorder-altered texture, which are comparable in magnitude, obscure it. However, the small (single-mT) stray fields around the edges of the disc suggest a curling magnetization, consistent with a vortex, as opposed to one tangential to the edges (see Supplementary Materials).

In both our simulations and measurements of the disc dynamics, we find that the pattern of AC stray fields generated by the vortex's azimuthal magnon mode are substantially easier to identify than the static stray field. The FWHM of the $\ket{0}\rightarrow\ket{+1}$ ODMR peak from our measurements of the whole disc can be seen in Fig.~\ref{fig:2PeakODMRDisc}(B), and the results of a zoomed-in, higher resolution scan around the vortex core (dashed box in Fig.~\ref{fig:2PeakODMRDisc}(B)) can be seen in Fig.~\ref{fig:2PeakODMRDisc}(C). In both simulations (Fig.~\ref{fig:2PeakODMRDisc}(E-F)) and experiment, find an anisotropic, extremely large MW field (FWHM) localized to a lobed structure near the core, with extended regions of smaller enhancement radiating outwards. Unlike with the previous wall mode measurements, we also see the anisotropy in these azimuthal mode measurements replicated in our simulations - this improvement is the result of adding grains to our simulations, and single-crystal simulations show a more symmetric result (see Supplementary Materials). The spatially extended field enhancement we see in both experiment and polycrystalline simulations, which don't appear in single crystal simulations (see Supplementary Materials), could be the result of magnon scattering, localized texture dynamics, or flux channeling at the grain edges, which presents a potential application of highly disordered permalloy in high frequency electronic applications\cite{rable_flux_2025}. These results, which have a resolution of approximately 56 nm (see Supplementary Materials), allow us to directly image the azimuthal mode near the core with 5$\times$ higher spatial resolution than the diffraction limited of optical measurements utilizing 515-532 nm light in air. To the best of our knowledge, this is the highest resolution experimental image of one of these modes reported in the literature\cite{park_interactions_2005,buess_excitations_2005}. 

\begin{figure} 
	\centering
	\includegraphics[width=0.6\textwidth]{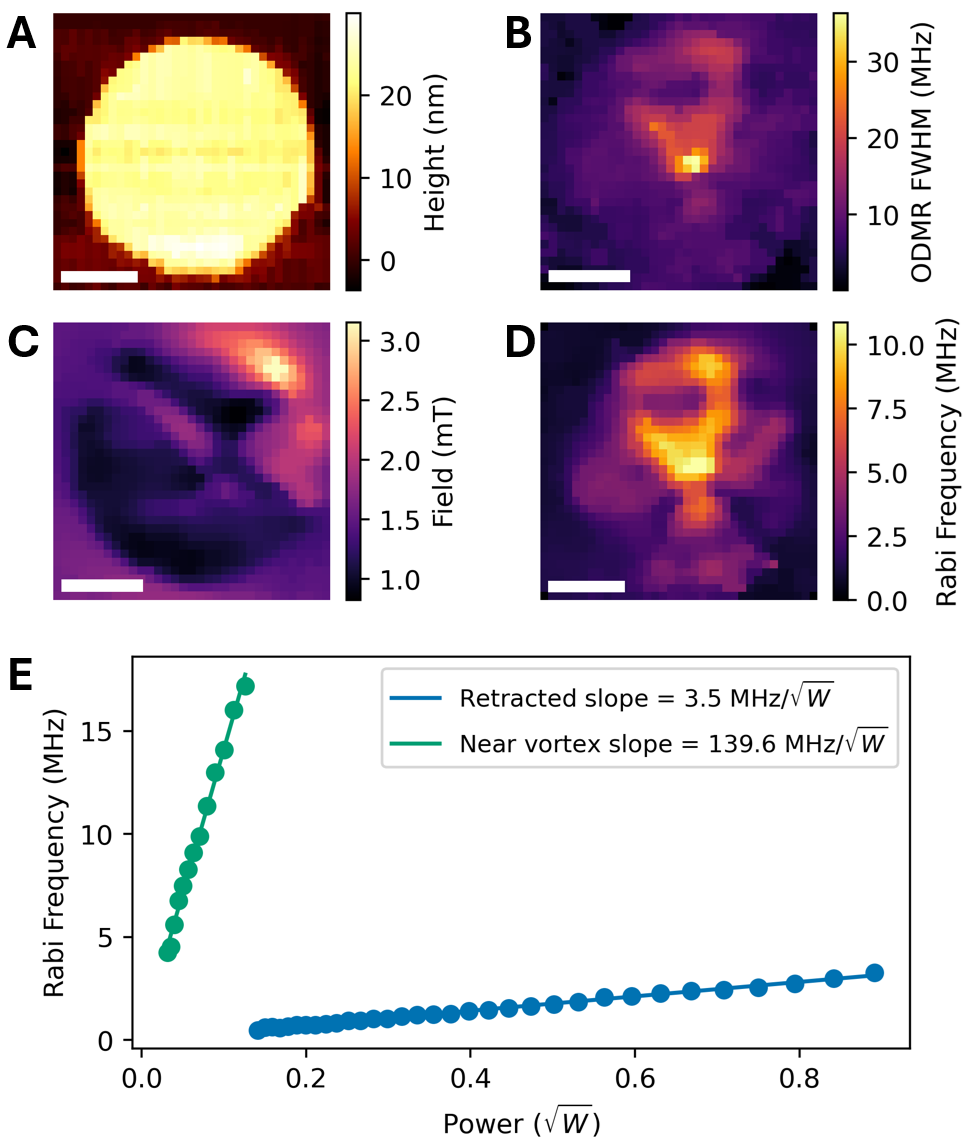} 
	\caption{\textbf{Rabi oscillation measurement results on a disordered, $6 \mu m$ disc at 300 nm lift.} Scale bars are 2 \textmu m.
		(A) Topography of the permalloy disc. (B) ODMR measurement results on the permalloy disc giving the approximate on-axis static field. (C) FWHM results extracted from the ODMR measurement shown in (B), showing qualitative variation in MW field strength. (D) Rabi oscillation scan results showing the quantitative variation in microwave field strength across the disc. These results correspond well with our ODMR FWHM. (E) Power dependence of Rabi oscillations measured at a single point near the vortex core and with the tip retracted off the sample. We find a 40$\times$ increase in MW power in the near-core measurement over the retracted measurement. (F) Individual Rabi oscillation measurements from the highest applied MW powers in (E) showing a large increase in Rabi frequency and decrease in contrast near the vortex compared to the retracted tip.}
	\label{fig:RabiDisc} 
\end{figure}

Our Rabi oscillation measurements in Fig.~\ref{fig:RabiDisc}, taken at a 300 nm lift height, demonstrate quantitative measurements of these azimuthal-mode generated fields, augmenting our results in Fig.~\ref{fig:2PeakODMRDisc}. Similar to the square results, we see a lower-resolution image of the ODMR results due to the lift height in Fig.~\ref{fig:RabiDisc}(B-C), and a quantitative map of the results in Fig.~\ref{fig:RabiDisc}(D). Here, we again see an anisotropic field emanating from the core, maintaining up to a 10$\times$ increase in Rabi frequency over our applied field despite the large NV-sample separation. Additionally, the reduced resolution makes our measurements more closely reflect existing optical measurements of vortex azimuthal magnon modes\cite{neudecker_modal_2006}. Measurements of the microwave enhancement produced by the vortex azimuthal mode, shown in Fig.~\ref{fig:RabiDisc}(E), show a 40$\times$ increase in the MW field amplitude near the vortex compared to the retracted case, similar to the greatest previously reported enhancement using a fixed NV center at 20 nm NV-sample separations\cite{wolf_strong_2017} on a 40 nm thick feature. Our similar coupling strength, despite the greater separation and thinner ferromagnet, is likely because we directly drive a vortex magnon mode - in the previous works\cite{Wolf2016,wolf_strong_2017}, the authors report coupling the NV to off-resonant vortex dynamics and expected much higher frequency ($>$4GHz) magnon modes. The magnitude of the Rabi oscillation enhancement we see corresponds to orders of magnitude reductions in the needed applied microwave power to achieve high Rabi frequencies, potentially leading to more efficient qubit driving with significantly less heating. 

\begin{figure} 
	\centering
	\includegraphics[]{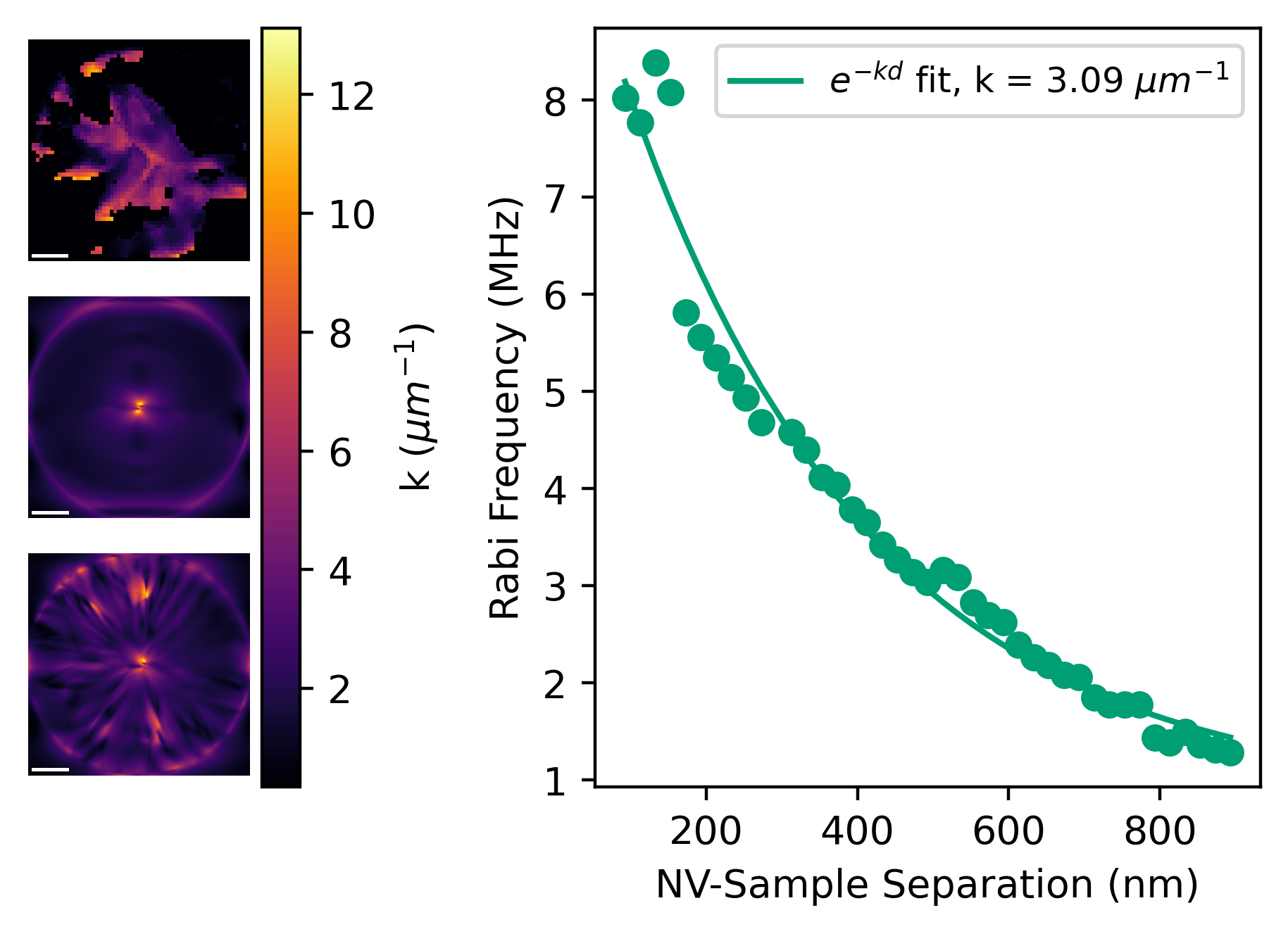} 
	\caption{\textbf{Evanescent decay of the MW fields produced by the disc.} Scale bars are 1 \textmu m.
		(A) Spatial variation in decay constants extracted from the height dependent exponential fit at each XY point in our ODMR scans (see Supplementary Materials). Points where the fit was poor are set to 0. (B) Spatial variation in decay constants extracted from the height dependent exponential fit at each point in space above an ideal, simulated permalloy disc. (C) Spatial variation in decay constants extracted from the height dependent exponential fit at each point in space above a polycrystalline, simulated permalloy disc with 1 \textmu m grains. This result matches our experiment better than the results in (B), showing that disorder significantly alters the decay of evanescent waves produced by the disc. (D) Height dependence of Rabi oscillations measured at single point near the vortex core. The result decays exponentially with a constant of 3.09 $\mu m^{-1}$.}
	\label{fig:K} 
\end{figure}

Next, we characterized the height dependence of the disc's microwave stray field utilizing both a single-point Rabi oscillation measurement near the vortex core (Fig.~\ref{fig:K}(D)), a series of ODMR measurements performed at different lift heights (Fig.~\ref{fig:K}(A)), and micromagnetic simulations (Figs.~\ref{fig:K}(B-C)). From our Rabi oscillation measurement near, but not at, the core, we find $k = 3.09 \mu m^{-1}$ ($323 nm$ decay length) , a common value in Fig.~\ref{fig:K}. This decay length, which is much longer than those of the square's wall mode, is a result of the greater spatial extent of the azimuthal mode, which is not confined to a domain wall. The stark contrast between the two systems opens new possibilities for qubit-magnon transduction. Because the cooperativity of such a transducer relies on the strength of the magnon stray field at the qubit location\cite{bejarano_parametric_2024}, future devices could be designed to optimize coupling at the necessary qubit-sample separation, which is set by material constraints and the need to reduce qubit exposure to magnetic noise\cite{Purser2020,Simon2022}. Additionally, unlike with the square, these comparatively long evanescent wave decay lengths make ODMR-based mapping more feasible - in these measurements, we perform a series of ODMR measurements at different tip lift heights and apply our exponential fit to the FWHM at each xy point in our maps. In these measurements, shown in Fig.~\ref{fig:K}(A-C), we find a substantial variation in field decay. In both of our ODMR measurements and simulations in Fig.~\ref{fig:K}(A-C), the amplitudes of the evanescent field decays have some dependence on their proximity to the vortex core, which is especially strong for the ideal case in Fig.~\ref{fig:K}(B), but this correlation becomes weaker when comparing to our experimental and grainy simulated decays in Figs.~\ref{fig:K}(A) and (C), respectively. While previous reports used the evanescent decay of magnon-generated microwave fields to identify plane waves with distinct wave vectors\cite{Simon2022}, here, both the mode geometry and granularity of our sample obfuscate such analysis, instead yielding a location dependent effective wave vector.

\subsection*{Discussion}

Overall, in this work, we have demonstrated that one can utilize SNVM to characterize, in depth, the MW stray field generated by two separate vortex modes - a wall mode in a square feature and an azimuthal magnon mode in a disc feature - with nanoscale resolution. These resolutions exceed that of conventional, diffraction-limited optical techniques by approximately a factor of 5, and, using recently reported methods to optimize NV-sample separation, could eventually reach an order of magnitude improvement, achieving a resolution comparable to that of X-ray microscopy techniques\cite{xu_minimizing_2025}. Unlike X-ray transmission microscopy techniques, such as TR-STXM, SNVM is also substrate-agnostic, allowing one to measure materials grown epitaxially on X-ray opaque materials, and does not require a beamline, allowing for greater measurement throughput.  In addition to imaging of these features' MW dynamics, we also demonstrated that it could be used to achieve up to a 40$\times$ increase in Rabi oscillation frequency of a qubit, and measured the decay of the evanescent waves. These results could prove useful in both the characterization of magnonic devices and the optimization of the many qubit-ferromagnet hybrid devices, such as transducers, that have been proposed, providing applications of this measurement technique to multiple fields. Future works could expand the applicable frequency range of this technique utilizing quantum frequency mixing\cite{karlson_quantum_2024, yin_high-resolution_2025, hu_nonlinear_2024} or other heterodyne sensing techniques\cite{meinel_heterodyne_2021}, enabling  characterization of the full breadth of not only linear and nonlinear vortex dynamics, but their complex interactions with magnons. 


\clearpage 

%
\bibliography{science_template} 
\bibliographystyle{sciencemag}

%
%
%
%
%
%


\section*{Acknowledgments}

\paragraph*{Funding:}
S.K. and A.B. acknowledge support provided by the National Science Foundation through the ExpandQISE award No. 2329067 and the Massachusetts Technology Collaborative through award number MTC-22032.
N.S. acknowledges support from the U.S. Department of Energy Office of Science National Quantum Information Science Research Centers (Q-NEXT).
\paragraph*{Author contributions:}
J.R. fabricated samples with J.D., performed the experimental measurements, ran the micromagnetic simulations, and analyzed the data with P.S. P.S., N.S., A.B., and S.K supervised the project.
\paragraph*{Competing interests:}
The authors have no competing interests to report. 
\paragraph*{Data and materials availability:}
All data needed to evaluate the conclusions in the paper are present in the paper and/or the Supplementary Materials.



\subsection*{Supplementary materials}
Materials and Methods\\
Supplementary Text\\
Figs. S1 to S13\\
References \textit{(30-\arabic{enumiv})}\\ 


\newpage


\renewcommand{\thefigure}{S\arabic{figure}}
\renewcommand{\thetable}{S\arabic{table}}
\renewcommand{\theequation}{S\arabic{equation}}
\renewcommand{\thepage}{S\arabic{page}}
\setcounter{figure}{0}
\setcounter{table}{0}
\setcounter{equation}{0}
\setcounter{page}{1} 


\begin{center}
\section*{Supplementary Materials for\\ \scititle}

	Jeffrey~Rable$^{\ast}$,
	Jyotirmay~Dwivedi,
	Nitin~Samarth, Paul~Stevenson, Arun~Bansil, Swastik~Kar$^{\ast}$\\
	\small$^\ast$Corresponding author. Email: j.rable@northeastern.edu, s.kar@northeastern.edu\\

\end{center}

\subsubsection*{This PDF file includes:}
Materials and Methods\\
Supplementary Text\\
Figures S1 to S13\\

\newpage


\subsection*{Materials and Methods}
\subsubsection*{Experimental Details}

We perform NV magnetometry measurements at room temperature using a commercial scanning system (Qnami ProteusQ), which combines a confocal microscope with an atomic force microscope, and an Akiyama probe with a diamond cantilever containing a single NV-center (Qnami Quantilever MX or MX+). Microwaves are applied using a wire loop that is manually positioned near the tip, and a static magnetic bias field is applied using a ring-shaped permanent magnet positioned around the objective lens above the sample. This applies an approximately 2.1 mT (disc) or 4.3 mT (square) field out of plane (OOP), and sets limits on the largest negative static fields we can reliably measure in our measurements - a Zeeman shift that pushes our NV transition peaks past the zero-field splitting will appear to `wrap' back around. 

ODMR measurements are performed by constantly illuminating the NV center with a 515 nm laser and measuring the emitted photoluminescence while sweeping the applied microwave frequency around the 2.87 GHz zero-field splitting of the NV center, with frequencies ranging from 2.5 to 3.3 GHz depending on the measurement. 

Rabi oscillation measurements are performed using the transition frequencies measured from a previous ODMR measurement at the same point. In these measurements, the NV spin is polarized into the $\ket{0}$ state via illumination with a 515 nm laser. Next, the laser is turned off and a MW field is applied at the transition frequency for a period of time, $\tau$, which is swept. Finally, readout of the NV spin is performed by turning the laser back on and measuring the NV photoluminescence. An additional reference measurement is performed after the NV spin has been repolarized into the $\ket{0}$ state, and the final result is the signal divided by the reference. This reference adjusts for any erroneous decrease in measured NV photoluminescence during the measurement.

Scans of ODMR are performed in two steps. First, the sample topography is measured either at each point (for a normal scan) or across a whole line (for a lifted scan). Next, an ODMR measurement is performed at each point to find the $\ket{0} \rightarrow \ket{1}$ transition or both the $\ket{0} \rightarrow \ket{1}$ and $\ket{0} \rightarrow \ket{-1}$ transition frequencies at the desired height. 

Rabi oscillation scans are performed in four steps. First, the sample topopgraphy is measured via a line scan, after which the tip is lifted 125 nm (square) or 300 nm (disc). Next, a course frequency-step ODMR measurement is performed to approximately find the NV $\ket{0} \rightarrow \ket{1}$ transition frequency. A finer ODMR measurement is then performed around the measured transition frequency at a lower microwave power to find a more precise value. Finally, a Rabi oscillation measurement is performed. 

Single point Rabi oscillation height (power) scans are performed similarly to the Rabi oscillation XY scans, using a course ODMR, fine ODMR, and Rabi oscillation measurement at each height (power). For the height scans, these changes are intended to account for changes in field as we change the NV-sample distance. For power scans, these additional ODMR scans are intended to account for drift - because the field gradients in close proximity to the vortex are large\cite{Wolf2016}, even nm-scale tip drifts could lead to pulse errors, skewing our measurements. 

\begin{figure} 
	\centering
	\includegraphics[width=\textwidth]{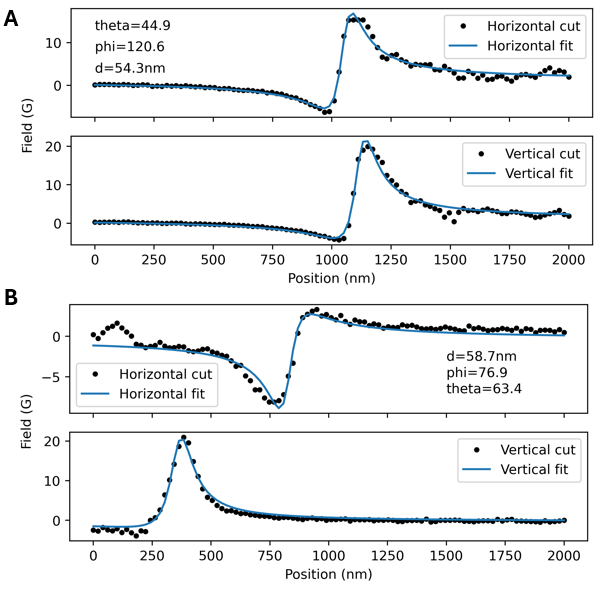} 
	\caption{\textbf{Calibration data taken over the edges of a multilayer sample with perpendicular magnetic anisotropy}.  (A) Calibration results from the tip used in the 1 \textmu m square measurements. (B) Calibration results for the tip used in the 6 \textmu m disc ODMR measurements. }
	\label{supfig:Calibration} 
\end{figure}

NV-sample distance and NV orientation are calibrated as outlined by Tetienne \textit{et. al.}\cite{Tetienne2015} using a standard sample composed of Ta(2 nm)/MgO/CoFeB(0.9 nm)/Ta(5 nm) with perpendicular magnetic anisotropy. The results for the tip used in all the measurements on the 1 \textmu m square feature can be seen in Fig.~\ref{supfig:Calibration}(A) and the results for the tip used in the disc ODMR measurements can be seen in Fig.~\ref{supfig:Calibration}(B).


\subsection*{Micromagnetic Simulations}
Micromagnetic simulations are performed using the Mumax3 software package, which solves for the magnetization of cubic cells using the Landau-Lifshitz-Gilbert equation\cite{Vansteenkiste2014}. Simulations are performed using 5x5x20 nm cells and standard permalloy material values - a saturation magnetization of $8.6 x 10^5\ A/m$, an exchange stiffness of $13\ pJ/m$, a uniaxial anistropy of $500\ J/m^3$ and a Gilbert damping constant of 0.02. In grainy simulations, Voronoi tesselation is used to create $1\ \mu m$ grains with a $\pm 5\%$ variance in saturation magnetization and uniaxial anisotropy. Exchange coupling between grains is also reduced by 5\%. The squares are modeled as a $20 nm$ thick, 1 \textmu m by 1 \textmu m side-length rectangle, and discs are modeled as a $6\ \mu m$ diameter, $20 nm$ thick cylinder, with both given an ideal vortex texture in the center. Then, they were allowed to relax in an out-of-plane static 4.3 mT (square) or 2.14 mT (disc) field, replicating our experimental conditions. Next, a 0.1 mT amplitude, 2.85 GHz external magnetic field is applied for 20 ns, with the cell magnetization sampled every 50 ps for 20 ns total runtime. This microwave field is applied at a 45$^{\circ}$ degree angle in the YZ plane of the simulation; this was chosen because we expect our antenna to produce a field at least partly in the plane of the sample. The stray field produced by the sample is found by applying a 2D Fast Fourier Transform (FFT) to the magnetization and finding the magnetic field in Fourier space, as detailed by Broadway \textit{et. al.}\cite{Broadway2020}. While you can find the stray field by creating extended simulation window of empty space in Mumax3, where the demagnetization field is the stray field, this FFT-based technique is significantly faster and can be performed at arbitrary distances from the sample. The magnitude of the AC stray field at 2.85 GHz is isolated using a Goertzel transform of the AC field - this captures the dynamics at the frequency of interest while discarding others excited in the disc, such as any potential harmonics or nonlinear dynamics\cite{heins_self-induced_2026}.

Additional micromagnetic simulations were also performed on the disc using a 2.7 to 3.05 GHz sinusoidal excitation, and the frequency of the response of the features was studied using a Gaussian pulse excitation with a 0.5 mT amplitude. 


\subsection*{Data Processing}
\subsubsection*{Drift Correction}
Notable drifts in our measurements occurred gradually and over a single axis. When drift occurred along the fast scan axis, drift correction was performed by assuming that the drift across each individual line was negligible compared to the total drift, and the data was re-centered using the topography data and the known shape of the feature. When drift occurred along the slow scan axis, drift correction was performed by comparing three separate points in the topography - the two edges and a piece of particulate near the center - to that of a fast reference measurement. The two areas between the three points were stretched or compressed to match the reference, assuming that the drift rate over the scan was approximately constant during those two portions of the scan.

\subsubsection*{Data Processing}
ODMR and Rabi oscillation scans have a median filter with a kernel of 3 applied using the SciPy python package. ODMR scan points with a poor Lorentzian fit (condition of covariance greater than $10^{10}$) are replaced using the median value of surrounding points prior to applying the median filter. Rabi oscillation scan points with a poor sinusoidal fit (condition of covariance greater than $5\times10^{14}$) are assigned a value corresponding to the maximum amplitude point extracted from an FFT of the data. The height scans in Fig.~\ref{fig:RabiDisc}(A) and Fig.~\ref{fig:RabiSquare}(A) has been leveled using a linear fit across the X-axis of the scan, and a median filter with a kernel of 7 applied along the slow scan direction to reduce streaking. Points in Fig.~\ref{fig:K}(A) with a negligible change in FWHM with height or a poor fit (condition of covariance of greater than $10^{10}$) replaced with 0. 


\subsection*{Supplementary Text}

\subsubsection*{Additional ODMR data}

\begin{figure} 
	\centering
	\includegraphics[width=\textwidth]{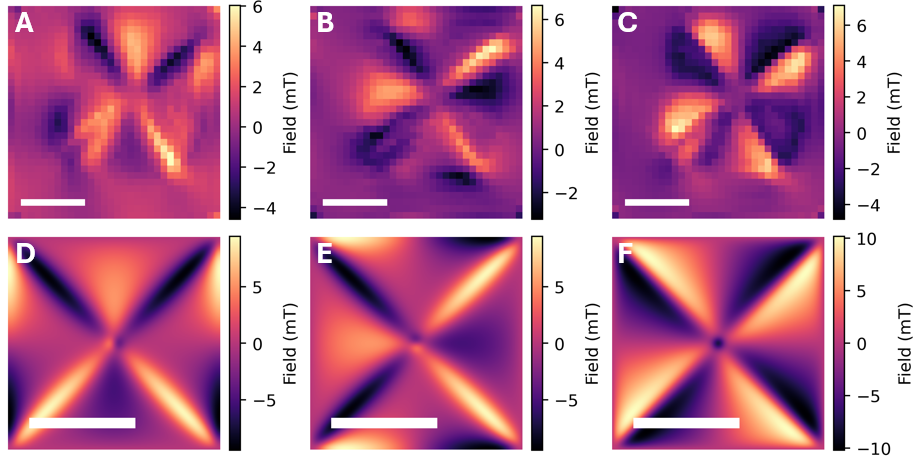} 
	\caption{\textbf{The cardinal components of the static stray field of the square}. Scale bars are 1\textmu m. (A-C) x, y, and z components, respectively, of square stray field in Fig.~\ref{fig:2PeakODMRSquare}(B). (D-F) x, y, and z components, respectively, of the simulated square stray field in Fig.~\ref{fig:2PeakODMRSquare}(E).}
	\label{supfig:StaticStrayFieldsCardinal} 
\end{figure}

Static stray field data from experimental measurements in Fig.~\ref{fig:2PeakODMRSquare} can also be decomposed into their components along the three cardinal directions in Fourier space using the known orientation of the tip - the results of this can be seen in Fig.~\ref{supfig:StaticStrayFieldsCardinal}(A-C). These fields show that the chirality of the vortex is counterclockwise, with a good match to the stray fields of a simulated counterclockwise vortex in Fig.~\ref{supfig:StaticStrayFieldsCardinal}(D-F). Were the vortex clockwise, the sign of the fields running from the core to the corners of the squares in Fig.~\ref{supfig:StaticStrayFieldsCardinal}(A-B) would be flipped. 

\begin{figure} 
	\centering
	\includegraphics[width=\textwidth]{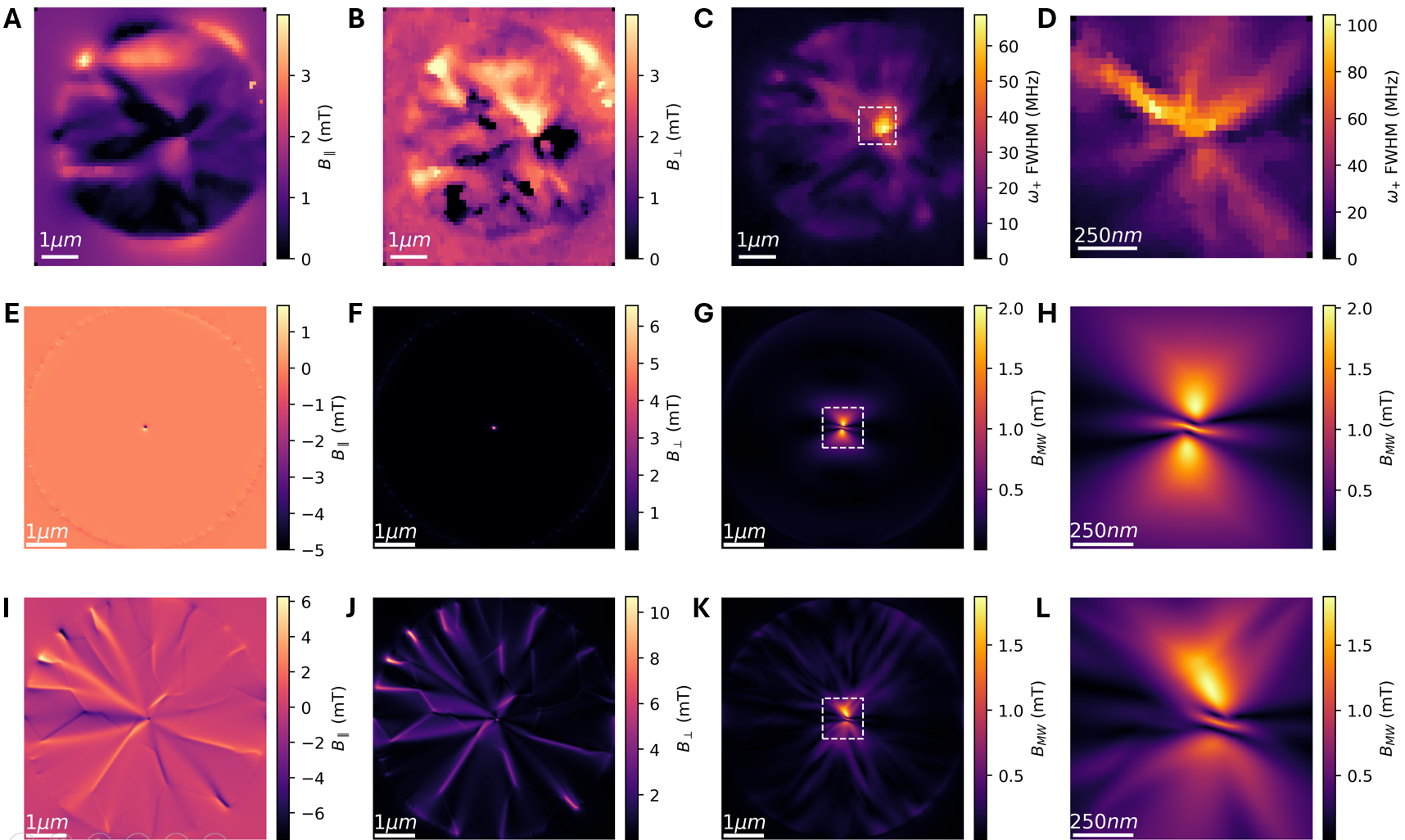} 
	\caption{\textbf{Extended version of Fig.~\ref{fig:2PeakODMRDisc} showing vector components of the static field and ideal disc simulations.}
		(A) Measurement of the static field parallel to the NV axis extracted from an ODMR measurement over the disc. (B) Measurement of the static field perpendicular to the NV axis extracted from an ODMR measurement over the disc. (C) FWHM extracted from the previous ODMR measurement, showing the position-dependent variation in MW field strength driven by the vortex. (D) FWHM extracted from a finer ODMR measurement in the $1 \mu m$ area around the vortex core (dashed box in (C)), showing the position-dependent variation in MW field strength driven by the vortex. (E-H) Simulated on-axis field, off-axis field, 2.85 GHz MW field under continuous driving across the whole disc and 2.85 GHz MW field under continuous driving near the vortex core, respectively. These simulations were performed on an ideal, single crystal disc. (I-L) Micromagnetic simulation results similar to (E-H), but performed on a polycrystalline disc composed of $1 \mu m$ grains.}
	\label{supfig:2PeakODMR} 
\end{figure}

Fig.~\ref{supfig:2PeakODMR} shows additional simulations on a single crystal vortex-containing disc, as well as the separated on and off-axis static stray field components measured by the NV center. The single crystal disc static stray fields (Fig.~\ref{supfig:2PeakODMR}(E-F)) do not match our measured results, but the dynamic stray fields near the vortex core (Fig.~\ref{supfig:2PeakODMR}(H)) still match fairly well.

\subsubsection*{Additional ODMR data - height dependence}

\begin{figure} 
	\centering
	\includegraphics[width=0.6\textwidth]{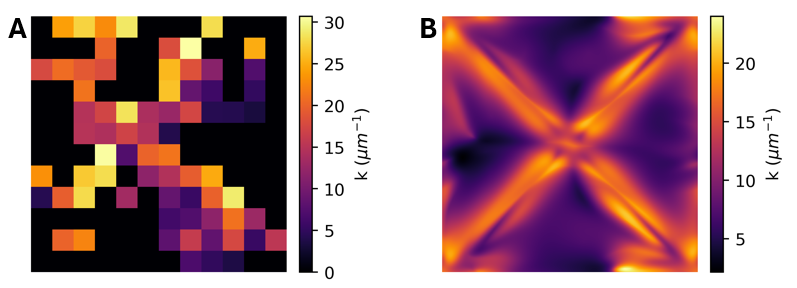} 
	\caption{\textbf{Estimated evanescent decay of MW stray fields over the square disc}
		 (A) Experimental data extracted from sweeps at different lift heights. (B) Simulated evanescent decay of MW from the disc.}
	\label{supfig:SquareK} 
\end{figure}

Additional lift height-swept ODMR scans over the square were performed, attempting to map the evanescent MW decay over the feature. These results, shown in Fig.~\ref{supfig:SquareK}(A), ultimately contained large errors because of the very rapid measured decays and our tip's fly height, which was comparable to the decay constant. Another potential source of error is the change in the FWHM as the microwave stray field decreases - once we are far enough away to leave the power broadened regime, the FWHM will drop off significantly faster. However, despite these issues, the results still somewhat qualitatively match simulated results in Fig.~\ref{supfig:SquareK}(B), which align with our single point measurements in Fig.~\ref{fig:RabiSquare}(F). 

\begin{figure} 
	\centering
	\includegraphics[width=0.3\textwidth]{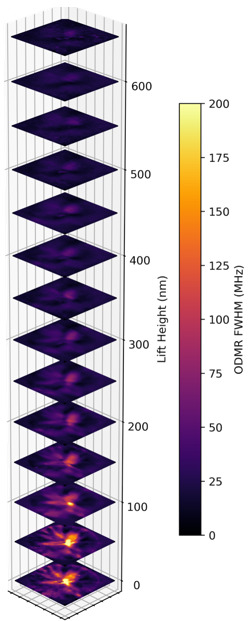} 
	\caption{\textbf{ODMR FWHM extracted from measurements at 0 nm lift height to 650 nm lift height over the disc feature.}
		 These FWHM results will correspond to the MW field amplitude linearly at higher powers and can be used to measure evanescent wave decay.}
	\label{supfig:HeightFWHM} 
\end{figure}

The data in Fig.~\ref{fig:K}(A) is the result of fitting the ODMR FWHM data in Fig.~\ref{supfig:HeightFWHM} at each XY point. This data shows what one would expect for  evanescent wave decay - the FWHM decreases rapidly with distance, and the features themselves blur out and become almost completely unresolvable by 650 nm lift height. 

\subsubsection*{Additional ODMR data - polarization analysis}

\begin{figure} 
	\centering
	\includegraphics[width=\textwidth]{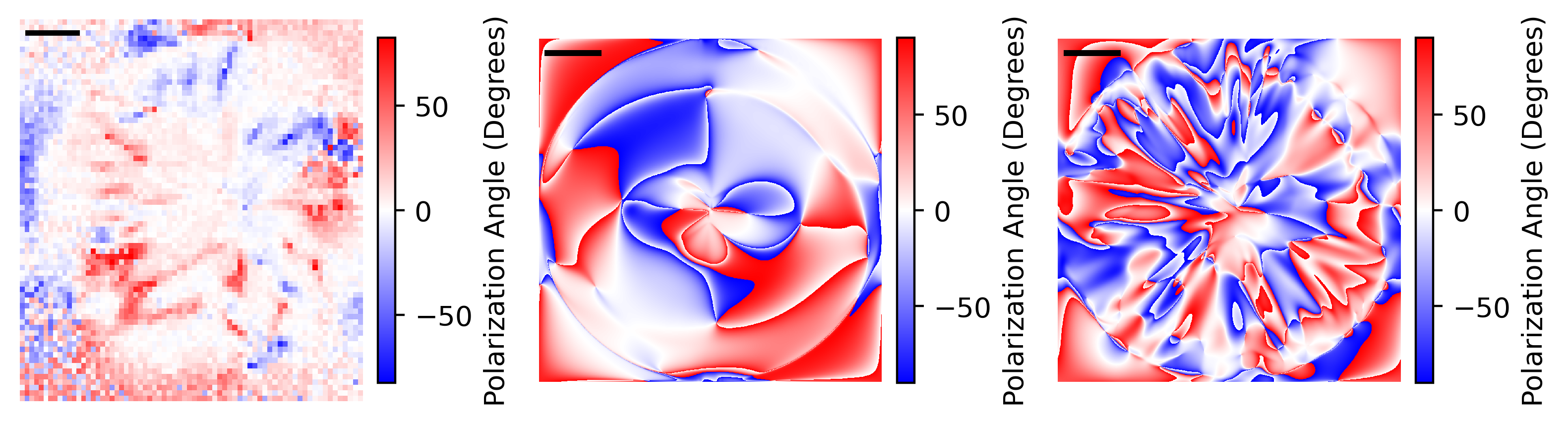} 
	\caption{\textbf{Polarization of MW field extracted from ODMR.} Scale bars are 1 \textmu m.
		(A) Polarization of the MW field extracted from the contrast of the two ODMR peaks measured in Fig.~\ref{fig:2PeakODMRDisc}. (B) Polarization of the MW field extracted from simulations of an ideal permalloy disc. (C) Polarization of the MW field extracted from simulations of a polycrystalline permalloy disc with 1 \textmu m grains.}
	\label{supfig:Polarization} 
\end{figure}

One significant potential advantage of SNVM over other imaging techniques is the ability to measure MW polarization, as right and left circularly polarized MW will preferentially excite the NV spin transitions such that:

\begin{equation}
    {\frac{C_1}{C_2}} = \sqrt{\frac{1-sin\phi}{1+sin\phi}}
    \label{eqn:polarization}
\end{equation}

where $C_1$ is the contrast of the $\ket{0}\rightarrow\ket{-1}$ transition, $C_2$ is the contrast of the $\ket{0}\rightarrow\ket{1}$ transition, and $\phi$ is the angle of the circular polarization of the microwaves, with 0 denoting linear polarization\cite{alegre_polarization-selective_2007,mrozek_circularly_2015,zheng_zero-field_2019}. This formula assumes that there is equal microwave power at both frequencies and no mixing between the ground state triplet spin states, and will therefore be an approximation for our measurements because of the variation in both MW and static stray fields across our sample. Our results, shown in Fig.~\ref{supfig:Polarization}(A), show a circularly symmetric variation in polarization along various `streaks' emitting outward from the center of the disc, with the most-circular polarizations lying closer to the edges of the disc, as expected given the aforementioned underestimation. This does not accurately match our simulations of an ideal disc in Fig.~\ref{supfig:Polarization}(B), but matches our simulations of a grainy disc in Fig.~\ref{supfig:Polarization}(C) significantly better.

\begin{figure} 
	\centering
	\includegraphics[width=\textwidth]{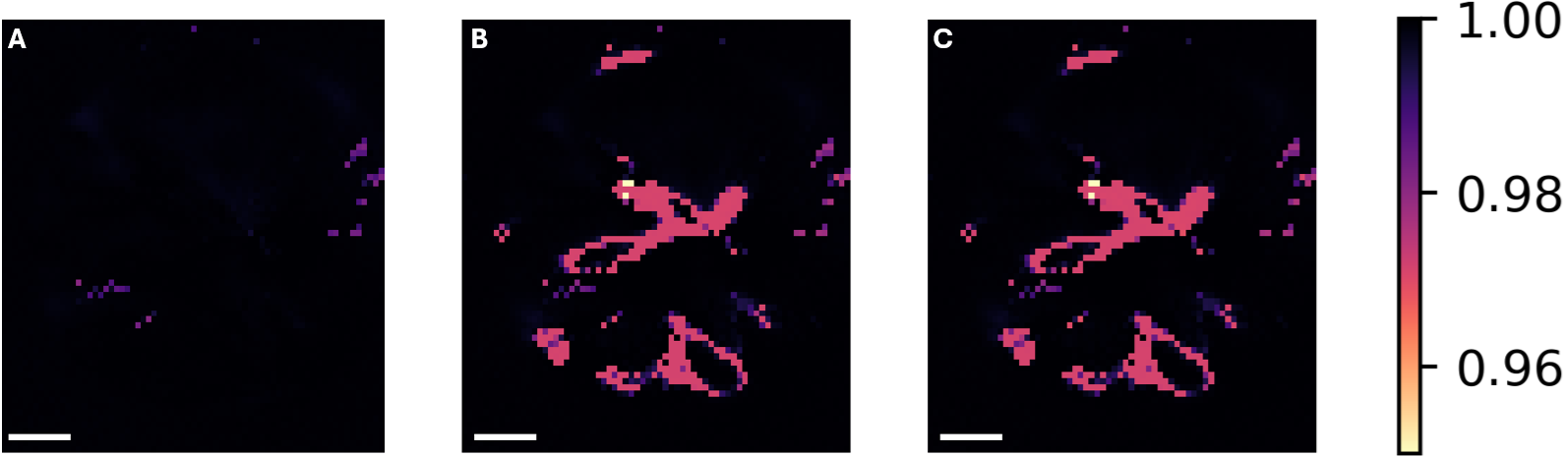} 
	\caption{\textbf{Similarity between eigenstates in the field over our permalloy disc and the zero-field eigenstates.}
		Scale bars are all $1 \mu m$. (A) Percent similarity between the new $\ket{0}$ eigenstate and the zero-field $\ket{0}$ eigenstate (B) Percent similarity between the new $\ket{-1}$ eigenstate and the zero-field $\ket{-1}$ eigenstate (C) Percent similarity between the new $\ket{+1}$ eigenstate and the zero-field $\ket{+1}$ eigenstate.}
	\label{supfig:eigen} 
\end{figure}

However, two separate confounding factors could affect our polarization measurements - spin-mixing, where the eigenstates of our original zero-field basis are no longer valid in the presence of off-axis magnetic fields, and variations in the dynamics of our disc at the two separate ODMR frequencies measured. To demonstrate that spin-mixing is not a major confounder in our results, we calculate the eigenstates at each point in our measurement using the built-in NV center ground state triplet model in the SimOS software package by Völker \textit{et. al.}\cite{volker_simos_2025}. The results, shown in Fig.~\ref{supfig:eigen}(A-C), show that the $\ket{0}$ state is nearly completely unaffected, and the $\ket{\pm1}$ states retain over 95\% of their zero-field character over most points in the measurement. The three brightest spots in Fig.~\ref{supfig:eigen}(B-C) near the center of the disc retain only 50\% of their zero-field character, but are likely erroneous and can be ignored. 

\begin{figure} 
	\centering
	\includegraphics[]{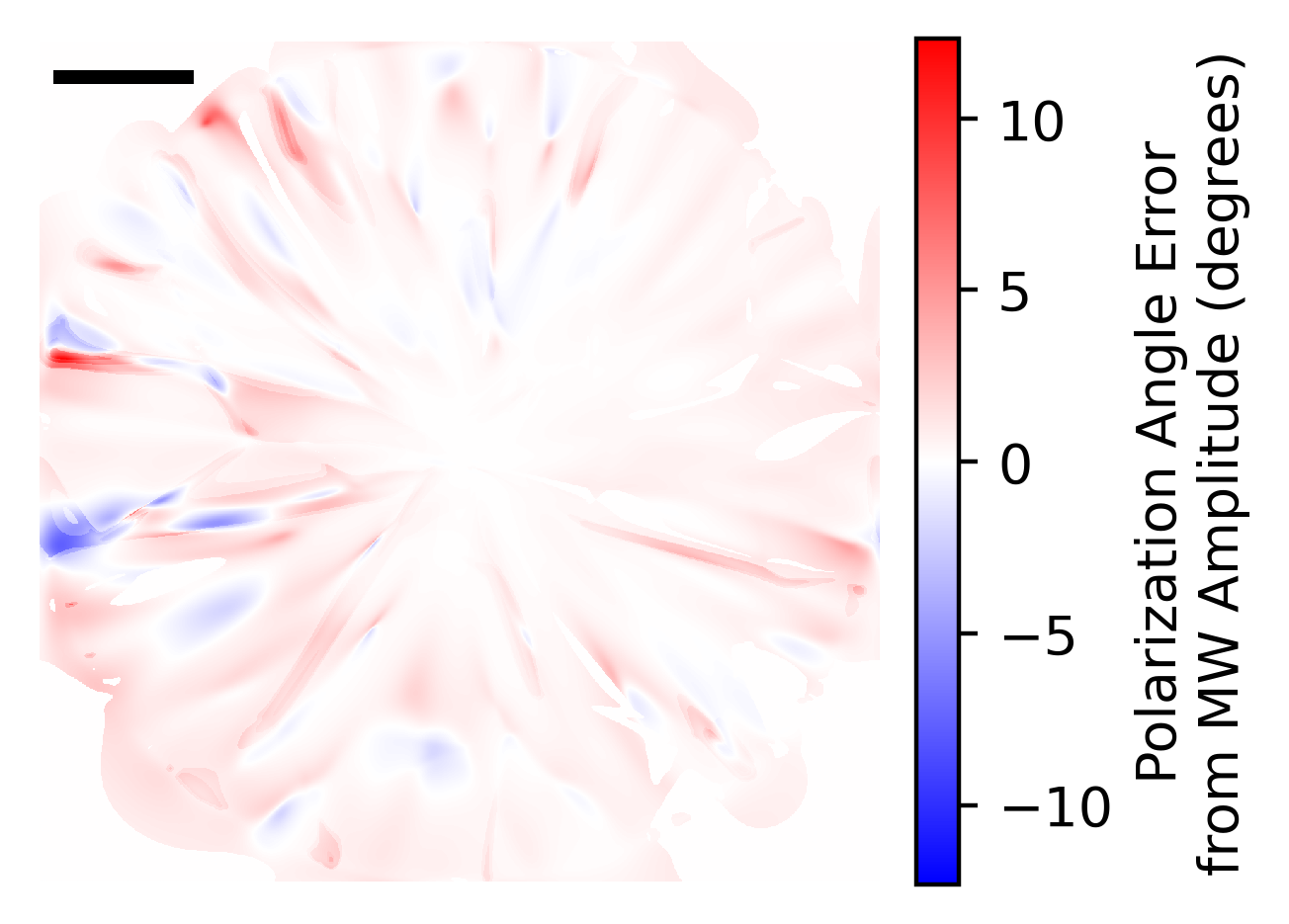} 
	\caption{\textbf{Calculated false 'polarization' values using the variable MW amplitudes at the two ODMR frequencies at each point in the simulated polycrystalline disc with 1 \textmu m grains.} Scale bar is 1 \textmu m. Because ODMR contrast is dependent on MW power as well as contrast, we utilized a series of simulations between 2.7 and 3.05 GHz to determine whether differences in MW power from the disc could be responsible for the results in our polarization analysis. These power variations are large enough to explain over one fifth of the polarization angles we see, suggesting that the analysis may contain large errors.}
	\label{supfig:FakePolarization} 
\end{figure}

We can estimate the impact of the other major potential confounding factor - differences in MW power at the two transition frequencies - by solving for the estimated power-induced contrast ratio and resulting contribution to the polarization calculation at each pair of measured frequencies. Simulations of the 6 \textmu m disc with 1 \textmu m grains are performed using 2.7, 2.75, 2.8, 2.85, 2.9, 2.95, 3, and 3.05 GHz sinusoidal MW drives. Using the static field data, the estimated NV transition frequencies $\omega_+$ and $\omega_-$ are calculated at each point and rounded to the nearest simulated drive frequency. Then, Rabi frequencies at the two transitions are calculated using the equation $\Omega_R = \gamma_{NV}B_{MW}/\sqrt{2}$, where $B_{MW}$ is the component of the disc stray field at the desired frequency plus the applied MW field. Estimated contrast ratios $\frac{C_1}{C_2}$ are calculated using the equation for continuous wave ODMR contrast from Dreau et al.\cite{Dreau2011}:

\begin{equation}
    C \approx \frac{\Omega_R^2}{\Omega_R^2+\Gamma_P^\infty\Gamma_C^\infty(\frac{s}{1+s})^2}
\end{equation}

where $\Gamma_P^\infty$ is 5 MHz, $\Gamma_C^\infty$ is 80 MHz, and s is a saturation parameter that can vary from about 0.01 to 0.3; we use 0.1 for our calculations. Constant factors were removed because they cancel when taking the ratio between the two calculated contrasts. This analysis results in Fig.~\ref{supfig:FakePolarization}, and suggests that variations in MW power emitted by the disc at different frequencies are expected to contribute up to nearly $12^{\circ}$ of error, suggesting an unfortunately large percentage of the results in Fig.~\ref{supfig:Polarization}(A) are not a result of polarization, but slightly different dynamical responses of the disk across the measured frequencies. Varying the saturation parameter predominately scales these values; for a saturation parameter of 0.05, the maximum, absolute error decreases to approximately 6 degrees, and for a  saturation of 0.2, the maximum, absolute error increases to approximately 24 degrees.

\subsubsection*{Additional Rabi oscillation analysis}

\begin{figure} 
	\centering
	\includegraphics[width=\textwidth]{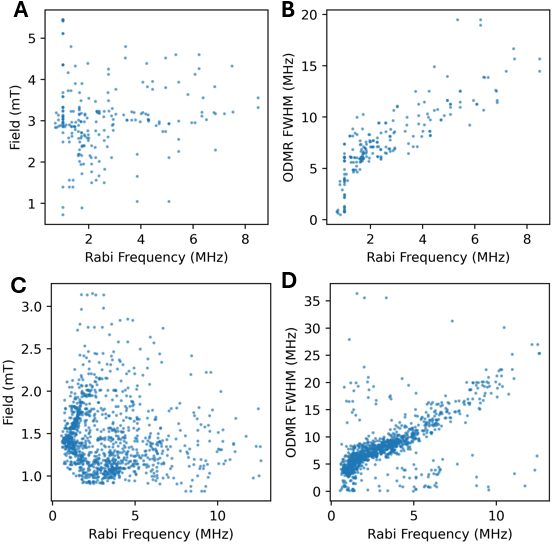} 
	\caption{\textbf{Scatter plots of data from fig.~\ref{fig:RabiSquare}(C-D) and fig.~\ref{fig:RabiDisc}(C-D)} (A) This data shows there is no strong correlation between stray field and Rabi frequency from the square. (B) This data shows that the FWHM and Rabi frequency correlate as expected - approximately linear in the power-broadening dominated regime, which is where the majority of data in Fig.~\ref{fig:2PeakODMRSquare} should lie. (C) This data shows there is no strong correlation between stray field and Rabi frequency from the disc. (D) Similar to (B), this data shows that the FWHM and Rabi frequency correlate as expected in Fig.~\ref{fig:2PeakODMRDisc}.}
	\label{supfig:RabiVsFWHM} 
\end{figure}

If our microwave antenna performance was extremely inconsistent over our applied field range, we would see a correlation between the measured ODMR frequency and microwave power, which would act as a source of error in our measurements. The scatter plot in Fig.~\ref{supfig:RabiVsFWHM}(A) and (C) shows there is no strong correlation between stray field (ODMR frequency) and Rabi frequency in Figs.~\ref{fig:RabiDisc} and ~\ref{fig:RabiSquare}, demonstrating that this is not a confounder in our measurements. To confirm that the Rabi frequency and ODMR FWHM are correlated as we expect, we plot the results from Figs.~\ref{fig:RabiSquare}(C-D) and ~\ref{fig:RabiDisc}(C-D) against each other as a scatter plot in Fig.~\ref{supfig:RabiVsFWHM}(B). These results show the expected linear dependence at high microwave driving powers, as expected. 


\subsubsection*{Field Calculation and Error Analysis}

The analytical solutions we use to calculate on and off-axis magnetic fields from ODMR transition frequencies in Fig.~\ref{fig:2PeakODMRSquare}(B-C), Fig.~\ref{fig:2PeakODMRDisc}(A), and Fig.~\ref{supfig:2PeakODMR}, derived by van der Sar \textit{et al.}\cite{VanderSar2015}, are:

\begin{equation}
    B_{||} = \frac{\sqrt{=(D+\omega_+-2\omega_-)(D+\omega_--2\omega_+)(D+\omega_-+\omega_+)}}{3\gamma\sqrt{3D}}
    \label{eqn:AnalyticalB}
\end{equation}

\begin{equation}
    B_{\perp} = \frac{\sqrt{-(2D-\omega_+-\omega_-)(2D+2\omega_--\omega_+)(2D-\omega_-+2\omega_+)}}{3\gamma\sqrt{3D}}
    \label{eqn:AnalyticalBPerp}
\end{equation}

where $\omega_\pm$ are the transition frequencies for the NV $\ket{0} \rightarrow \ket{\pm 1}$ transitions, $D_{gs}$ is the NV zero-field splitting frequency, typically 2.87 GHz, and $\gamma_{NV}$ is the NV center gyromagnetic ratio of 28.025 GHz/T. 
However, in order to reduce measurement time, many of our ODMR measurements only measured the $\omega_+$ transition frequency - this allows us to narrow the frequency parameter space we sweep, significantly speeding up measurements where we were only interested in the ODMR FWHM or the transition frequency, and not the exact static fields. For example, while measuring Rabi oscillations, where we only need a single transition frequency, or when trying to map the spatial variation in evanescent decays. Using a single transition frequency, the on-axis field can be estimated as:

\begin{equation}
    B_{||} \approx \gamma_{NV}|D_{gs} - \omega_{\pm}|
    \label{eqn:NaiveB}
\end{equation}
This approximation ignores the effects of off-axis fields, and will become more inaccurate as the off-axis field increases relative to the on-axis field. We can determine the error of these estimated magnetic fields with Eqn.~\ref{eqn:NaiveB} by comparing them to the analytical results from Fig.~\ref{fig:2PeakODMRDisc}(A-B). 

\begin{figure} 
	\centering
	\includegraphics[width=\textwidth]{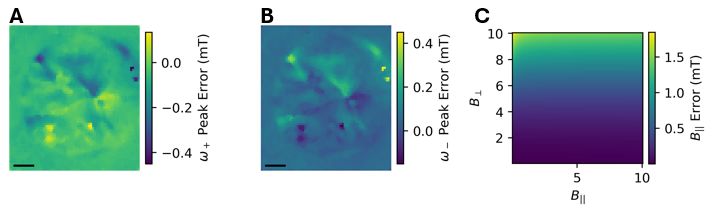} 
	\caption{\textbf{Errors from calculating on-axis magnetic fields using only a single ODMR peak for the disc feature.} Scale bars are 1 \textmu m. (A) Error in on-axis magnetic field measurements when calculated using only the $\omega_+$ peak. (B) Error in on-axis magnetic field measurements when calculated using only the $\omega_-$ peak. (C) Error in the on-axis magnetic field relative to on and off-axis magnetic fields up to 10 mT.}
	\label{supfig:errors} 
\end{figure}

Comparing these approximations to our analytical results using Eqn.~\ref{eqn:AnalyticalB} in Fig.~\ref{supfig:errors}(A-B), we find errors of up to approximately 0.45 mT from the single peak approximations for our data. Because off-axis fields shift both peaks to a higher frequency with almost equal amplitude in the low field limit\cite{welter_scanning_2022,beaver_optimizing_2024}, the average of these two results nearly gives the same result as Eqn.~\ref{eqn:AnalyticalB}, with under 0.03 mT error despite being a drastically simplified approximation. Fig.~\ref{supfig:errors}(C) shows the absolute error in calculated on-axis field, $B_{||}$, for on and off-axis fields up to 10 mT. The error peaks at 1.5 mT in an off-axis 10 mT field.

Based on these results, we can expect the error in Fig.~\ref{fig:RabiDisc}(B) and Fig.~\ref{fig:RabiSquare} to be relatively small.  

\subsubsection*{Additional micromagnetic simulations data}
\begin{figure} 
	\centering
	\includegraphics[width=0.6\textwidth]{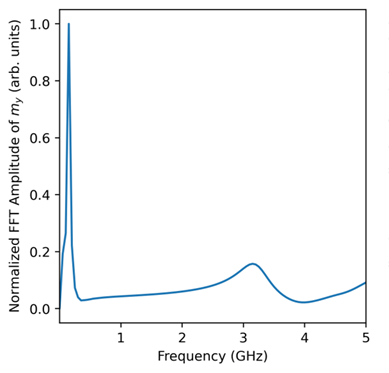} 
	\caption{\textbf{FFT of the y component of the reduced magnetization of the vortex-containing $1\mu m$ square under a Gaussian pulse.} Like the disc, these results show two primary modes - a low frequency gyrating mode and a broader wall mode near the NV center transition frequencies.}
	\label{supfig:fft_square} 
\end{figure}

\begin{figure} 
	\centering
	\includegraphics[width=0.6\textwidth]{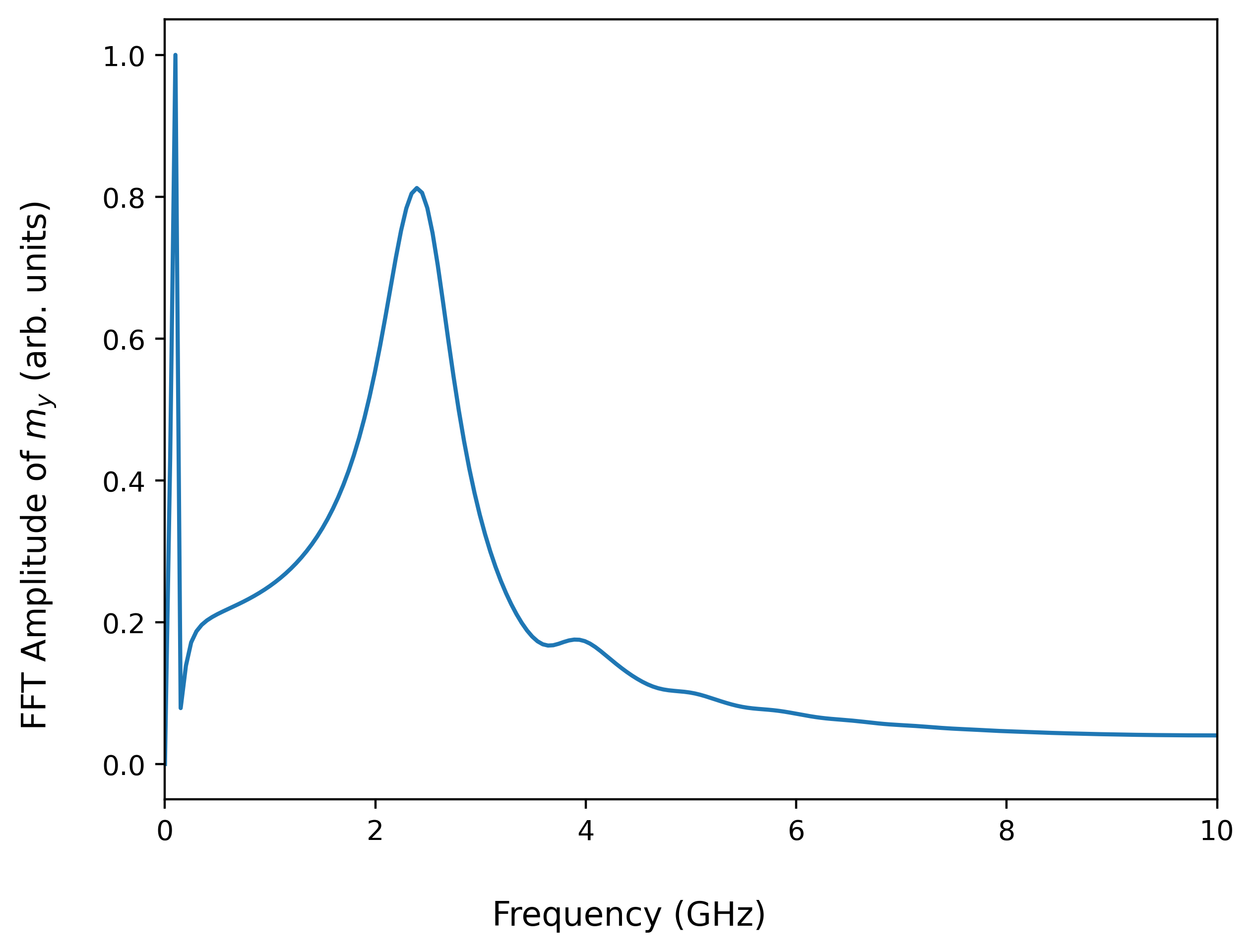} 
	\caption{\textbf{FFT of the y component of the reduced magnetization of the vortex-containing, $6 \mu m$ diameter disc under a Gaussian pulse.} These results show two primary modes - a low frequency, gyrating mode below 100 MHz, and a broad, azimuthal mode near the NV center transition frequencies.}
	\label{supfig:fft_disc} 
\end{figure}

The dynamics of the $1\mu m$ square's average magnetization under a Gaussian pulse can be seen in Fig.~\ref{supfig:fft_square}. Here, we can see two modes - a gyrating mode and a higher frequency mode, the latter of which overlaps with the NV transition frequencies. Similar results for the $6 \mu m$ disc can be seen in Fig.~\ref{supfig:fft_disc}.

\begin{figure} 
	\centering
	\includegraphics[width=\textwidth]{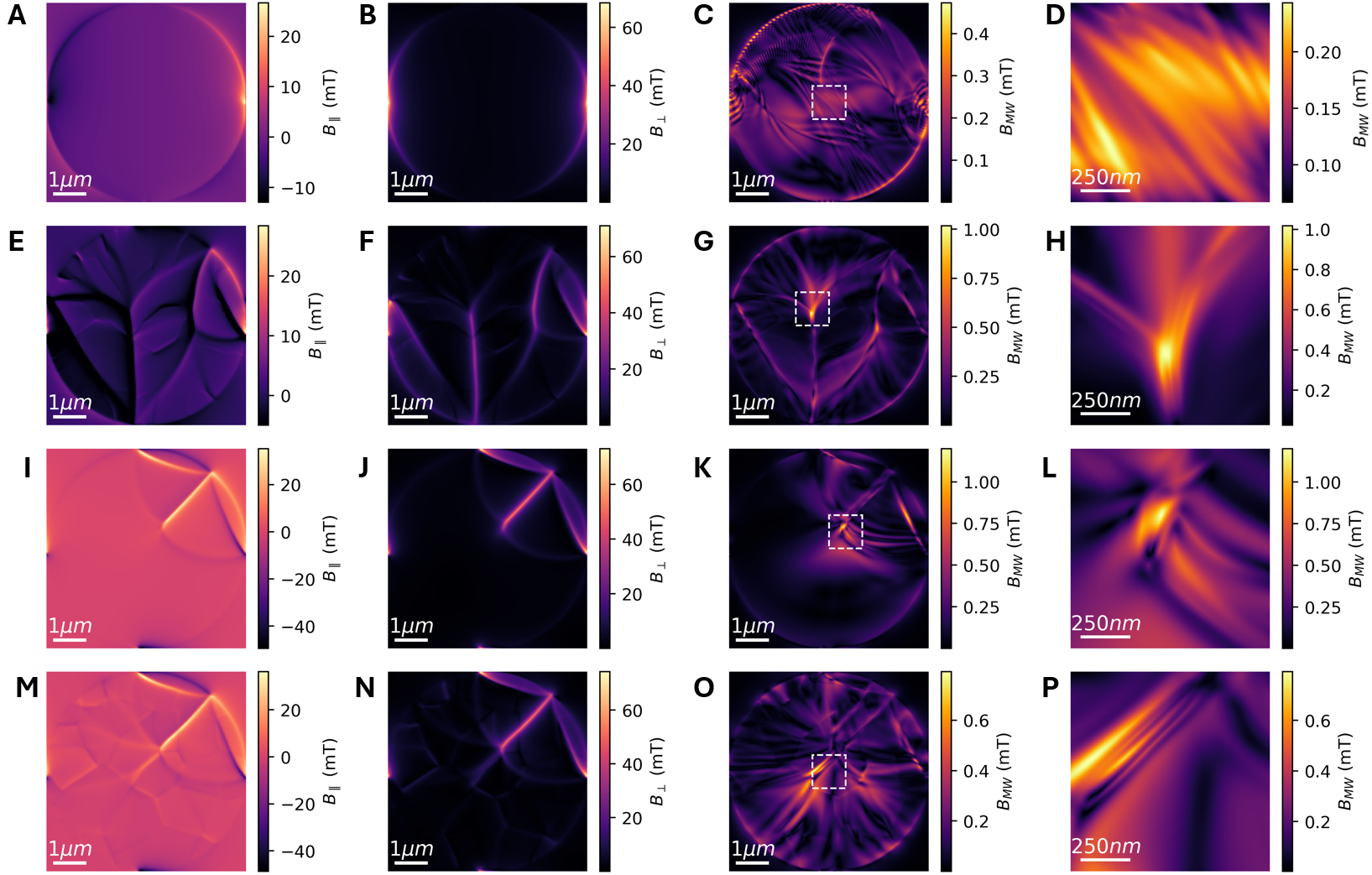} 
	\caption{\textbf{Simulated static and dynamic stray fields in alternative magnetization states}
		 (A-D) Static and 2.85 GHz stray fields under 2.85 GHz excitation from a single crystal permalloy disc initialized with a uniform magnetization along the X direction. (E-H) Static and 2.85 GHz stray fields under 2.85 GHz excitation from a polycrystalline  permalloy disc initialized with a uniform magnetization along the X direction. The grains are 1 \textmu m. (I-L) Static and 2.85 GHz stray fields under 2.85 GHz excitation from a single crystal permalloy disc initialized with a transverse domain wall in the center of the disc. The domain wall relaxed into a state where it extends from the center of the disc to the upper-right corner of the disc, with a more gradual rotation between the two states in the other areas of the disc. (M-P) Static and 2.85 GHz stray fields under 2.85 GHz excitation from a polycrystalline permalloy disc initialized with the relaxed magnetization state in (I-L).}
	\label{supfig:AltMag} 
\end{figure}

To confirm that other textures and dynamics could not be responsible for our results in the main text, we ran simulations for single crystal and polycrystalline permalloy discs with different initial magnetization in Fig.~\ref{supfig:AltMag}. Firstly, none of these results show comparable static stray fields to our experimental results - they are all multiple times larger and qualitatively different than the fields we measured. Secondly, we find that none of these configurations produces large enough microwave fields to explain the magnitude of our results in Fig.~\ref{fig:2PeakODMRDisc}(C-D), nor do they accurately reproduce the same signal shape we see in our experiments. In the saturated cases, where off-resonant flux channeling will dominate, we see an approximately 2$\times$ increase in MW field magnitude in the single crystal disc (Fig.~\ref{supfig:AltMag}(A-D)), and a 10$\times$ increase in MW field at the grain boundaries and textures in the polycrystalline disc (Fig.~\ref{supfig:AltMag}(E-H)), suggesting a potential application of highly disordered soft ferromagnets for localized flux channeling and microwave amplification. Similarly, in both the single and polycrystalline discs with a stable transverse domain wall (Fig.~\ref{supfig:AltMag}(I-P)), we find approximately 10$\times$ and 8$\times$ increases in microwave field at specific sites along the disc, showing that these magnetization configurations are off from our experiment by a factor of 4 to 20, whereas the vortex results were only off by a factor of 2. Furthermore, all our attempts at introducing a transverse domain wall in the polycrystalline disc simulations in Fig.~\ref{supfig:AltMag}(M-P) ultimately produced a vortex state when allowed to relax into the lowest energy state, suggesting they are the more energetically favorable texture in a polycrystalline disc. To circumvent this, we used the relaxed magnetization of the single crystal disc in Fig.~\ref{supfig:AltMag}(I-L) as our initial magnetization state for the polycrystalline simulations.

\end{document}